\documentclass{article}
\usepackage{ijcai25}
\usepackage{times}
\usepackage{soul}
\usepackage{url}
\usepackage{xurl} 
\usepackage[hidelinks]{hyperref}
\usepackage[utf8]{inputenc}
\usepackage[small]{caption}
\usepackage{graphicx}
\usepackage{amsmath}
\usepackage{amsthm}
\usepackage{booktabs}
\usepackage{algorithm}
\usepackage{algorithmic}
\usepackage{listings}
\usepackage{xcolor}
\usepackage{breakurl}
\usepackage{float}
\usepackage{booktabs}
\usepackage{array}
\usepackage{multirow}
\usepackage[switch]{lineno}

\usepackage[inline]{enumitem}
\setlist[itemize]{left=0pt, labelsep=0.5em, itemsep=0.3em}
\usepackage{titlesec}

\titlespacing*{\section}
  {0pt}   
  {0.5em}   
  {0.25em} 

\titlespacing*{\subsection}
  {0pt}
  {0.4em}
  {0.1em}

\titlespacing*{\paragraph}
  {0pt}    
  {0.1em}  
  {0.5em}  
  
\title{AI4Contracts: LLM \& RAG-Powered Encoding of Financial Derivative Contracts}

\author{
Maruf Ahmed Mridul$^1$\and
Ian Sloyan$^2$\and
Aparna Gupta$^1$\And
Oshani Seneviratne$^1$
\affiliations
$^1$Rensselaer Polytechnic Institute,  $^2$South Cardinal\\
\emails
mridum@rpi.edu,
ian@southcardinal.ie,
guptaa@rpi.edu, 
senevo@rpi.edu
}

\begin{document}

\maketitle

\begin{abstract}
Large Language Models (LLMs) and Retrieval-Augmented Generation (RAG) are reshaping how AI systems extract and organize information from unstructured text. A key challenge is designing AI methods that can incrementally extract, structure, and validate information while preserving hierarchical and contextual relationships. We introduce \emph{CDMizer}, a template-driven, LLM, and RAG-based framework for structured text transformation. By leveraging depth-based retrieval and hierarchical generation, \emph{CDMizer} ensures a controlled, modular process that aligns generated outputs with predefined schema. Its template-driven approach guarantees syntactic correctness, schema adherence, and improved scalability, addressing key limitations of direct generation methods. Additionally, we propose an LLM-powered evaluation framework to assess the completeness and accuracy of structured representations. Demonstrated in the transformation of Over-the-Counter (OTC) financial derivative contracts into the Common Domain Model (CDM), \emph{CDMizer} establishes a scalable foundation for AI-driven document understanding, structured synthesis, and automated validation in broader contexts.
\end{abstract}

\vspace{-1em}
\section{Introduction}

Advancements in Large Language Models (LLMs) and Retrieval-Augmented Generation (RAG) are transforming how AI systems extract and organize unstructured text. However, applying these techniques to long, hierarchical, and domain-specific documents remains challenging. AI methods must extract key information while generating structured outputs that align with predefined schema and preserve semantic integrity.  

We introduce \emph{CDMizer}, a template-driven, LLM, and RAG-based framework for structured text transformation. By leveraging depth-based retrieval and hierarchical generation, \emph{CDMizer} extracts and organizes unstructured contract text while ensuring syntactic correctness, schema adherence, and scalability. This process reduces manual overhead, enhances efficiency, and facilitates compliance in high-stakes financial environments. We also propose an \emph{LLM-powered evaluation framework} to assess the structured output quality.  

Structured representations are critical in financial contracts, where unstructured text complicates automation and compliance.
CDM, a standardized, machine-readable, and machine-executable model, represents financial products, trades in those products, and the lifecycle events of those trades~\cite{finos_cdm}. International Swaps and Derivatives Association (ISDA), the organization driving the development of CDM at Fintech Open Source Foundation (FINOS) (along with International Capital Market Association and International Securities Lending Association), provides legal standards for trillions of dollars of notional contracts~\cite{isda}. However, these contracts often require manual processing due to their unstructured nature and complex terms. The challenge is greater for non-centrally cleared OTC derivatives, where counterparties handle collateral, settlement, and disputes bilaterally. While CDM offers a standardized schema, converting complex legal agreements into structured representations remains difficult due to domain-specific clauses and multilayered structures.

By structuring contract data into CDM-aligned outputs, \emph{CDMizer} automates and standardizes OTC derivatives processing, bridging the gap between unstructured financial agreements and machine-readable contract frameworks.



\paragraph{Contributions} of this paper include: (1) \textbf{Baseline Conversion Method:} An LLM and RAG pipeline for generating CDM representations from unstructured derivative contracts, serving as a feasibility benchmark for AI-driven solutions, (2) \textbf{Template Creation Framework:} A deterministic method for creating minimal yet comprehensive OTC derivatives templates, retaining only essential CDM fields, (3) \textbf{\emph{CDMizer}:} A structured method that builds on the templates, integrating LLMs and RAG for scalable contract-to-CDM conversion while ensuring syntactic correctness and schema adherence, and (4) \textbf{Evaluation Framework:} An LLM-driven approach to assess the semantic completeness and correctness of generated CDM representations at scale.

\vspace{-0.1em}
\section{Background}
\subsection{Common Domain Model (CDM)}
FINOS CDM, originally developed by ISDA, is a standardized, machine-readable, and machine-executable blueprint for how financial products are traded and managed across the transaction lifecycle. It enhances efficiency, interoperability, and transparency in financial markets by providing a single digital processing standard, reducing reconciliation needs, accelerating financial technology innovation, and enabling consistent regulatory reporting. Built on key design principles—normalization, composability, industry format mapping, embedded logic, and modularization—CDM ensures a structured, adaptable framework for trade processing~\cite{cdm_purpose}.
The CDM model, maintained in the FINOS CDM repository~\cite{finos-cdm-repository}, is the central resource for defining the structure and semantics of OTC derivatives and other financial contracts. This repository also provides numerous example contracts in CDM-compliant JSON and schema definitions that outline validation rules and structural guidelines to ensure consistency.
While CDM applies to a wide set of financial products, this work focuses on OTC derivatives. 

\subsection{OTC Financial Derivative Contracts}
Financial derivatives are financial contracts defined as a derived contract on another underlying financial contract, instrument, index, or reference rate. These contracts allow parties to speculate on or hedge against changes in value or volatility of the underlying financial contract or asset, which may be an equity, commodity, bond, interest rate, currency, or other financial instrument. Derivatives, especially the OTC ones, have a large variety, including forwards, swaps, and options.
OTC derivatives are bilaterally negotiated financial contracts between two counterparties without being listed or traded on an exchange. Unlike exchange-traded derivatives, which are standardized, OTC derivatives are customized to meet counterparties' specific risk management needs. These customized contracts are constructed on a wide range of asset classes, such as fixed income, equity, foreign exchange, commodity, credit instruments, etc, and are often used for hedging risk or speculating on market movement and volatility.

\section{Related Work}

Recent advances in reg-tech and financial automation have demonstrated the potential of LLMs and RAG in streamlining complex processes. While several studies emphasize standardized formats, few directly address converting unstructured OTC derivative contracts into structured representations. The following works highlight key developments and gaps in this domain.

\paragraph{Early Automated Frameworks for OTC Derivatives:}
~\cite{fries2018smart} proposed automating termination procedures in collateralized OTC transactions but relied on structured contract terms. Similarly, ~\cite{armitage2022trust} explored transforming the ISDA Master Agreement into a smart contract, emphasizing systematic structures but not AI-driven free-text conversion.

\paragraph{Toward Smart Contract Implementations:}
~\cite{oluwajebe2020smart} demonstrated smart contract automation for interest rate swaps (IRS), underscoring the need for standardized templates. Formal methods in ~\cite{clack2018temporal} reinforced the importance of clear semantics for legal enforceability. However, these works assume structured inputs, overlooking challenges in extracting structured data from unstructured legal text.

\paragraph{LLMs, Code Generation, and Interoperability:}
~\cite{kang2024using,van2023translating} showed how LLMs generate smart contracts from health insurance policy documents but did not address complex financial agreements. SolMover ~\cite{karanjai2024solmover} tackled blockchain contract translation, leveraging structured representations. Additionally, ~\cite{sorensen2024correct} highlighted AI-induced failures in smart contracts, stressing the need for precise definitions.

\paragraph{Decentralized Finance and Retrieval-Based Methods:}
~\cite{singh2024option} examined decentralized exchanges and how inconsistent contract specifications hinder large-scale adoption. ~\cite{vaithilingam2022expectation} emphasized human oversight in LLM-driven code generation, while REDCODER ~\cite{parvez2021retrieval} demonstrated how retrieval-augmented generation enhances coherence and correctness, making it relevant for contract standardization.

\paragraph{Evaluating Large-Scale AI-Driven Translation:}
A key challenge in contract standardization is evaluating AI-generated structured outputs. Many AI-driven translation efforts ~\cite{kang2024using,van2023translating,vaithilingam2022expectation,parvez2021retrieval,liu2024your,li2022competition} rely on test-based evaluations, which are impractical for complex financial contracts. While round-trip correctness (RTC) ~\cite{allamanis2024unsupervised} aids in code evaluation, it remains unsuitable for contract standardization due to legal nuances. Developing robust evaluation metrics remains an open challenge.

In summary, most existing solutions assume structured inputs or focus on narrow contract components, leaving the transformation of full-length OTC derivative contracts unresolved. This paper addresses this gap by introducing the CDMizer, and a robust evaluation framework.

\section{Data Collection and Synthesis}


The first step in generating CDM representations is acquiring the necessary data. However, real-world OTC derivative contracts are often proprietary and confidential, limiting access to diverse examples that capture the full complexity and variability of real agreements. This constraint poses a significant challenge, as synthetic samples may fail to reflect the nuanced language and edge cases present in actual contracts, hindering the development of robust automation solutions. Nevertheless, we were able to collect the following data:

\begin{itemize}[topsep=0pt, partopsep=0pt, itemsep=0pt, parsep=0pt, left=0pt]
    \item \textbf{Natural Language Contracts (Term Sheets):} Two contract examples from RBCCapitalMarkets~\cite{rbc-structuredrates-2024}
    and JPMorgan~\cite{jpm-usd-inr-irs}.
    \item \textbf{CDM Examples:} 858 structured CDM instances sourced from the FINOS CDM repository~\cite{finos-cdm-repository}.
\end{itemize}

Although the CDM repository provides structured JSON examples, natural language contract descriptions remain scarce. Notably, while CDM representations exist, corresponding real-world contract descriptions in natural language are not readily available. Since our objective is to convert natural language descriptions into CDM format, obtaining such descriptions is a prerequisite.

To address this gap, we leveraged an LLM to generate synthetic contract descriptions from existing CDM representations. To enhance the quality of these generated descriptions, we provided the two collected natural language examples as reference inputs. While synthetic descriptions may not fully capture the complexity of real-world contracts, they serve as an initial dataset for training and model development.

For experimentation purposes and convenience, we categorized the collected and generated data into six contract types: \textit{Interest Rate Swap}, \textit{Equity Swap}, \textit{Equity Option}, \textit{Commodity Option}, \textit{Foreign Exchange} Derivatives, and \textit{Credit Default Swap}. 
This classification was guided by the collected CDM examples, each of which explicitly specifies its contract type, reflecting the primary categories represented in the available structured data.

\section{Baseline Method}

The baseline method, illustrated in Figure~\ref{fig:workflow_1}, evaluates how well LLMs can generate CDM representations directly from OTC derivative contract descriptions. The process involves fine-tuning an LLM on contract-to-CDM mappings and integrating RAG for enhanced accuracy. The steps are as follows:

\begin{figure}[h]
    \centering
    \includegraphics[width=0.9\linewidth]{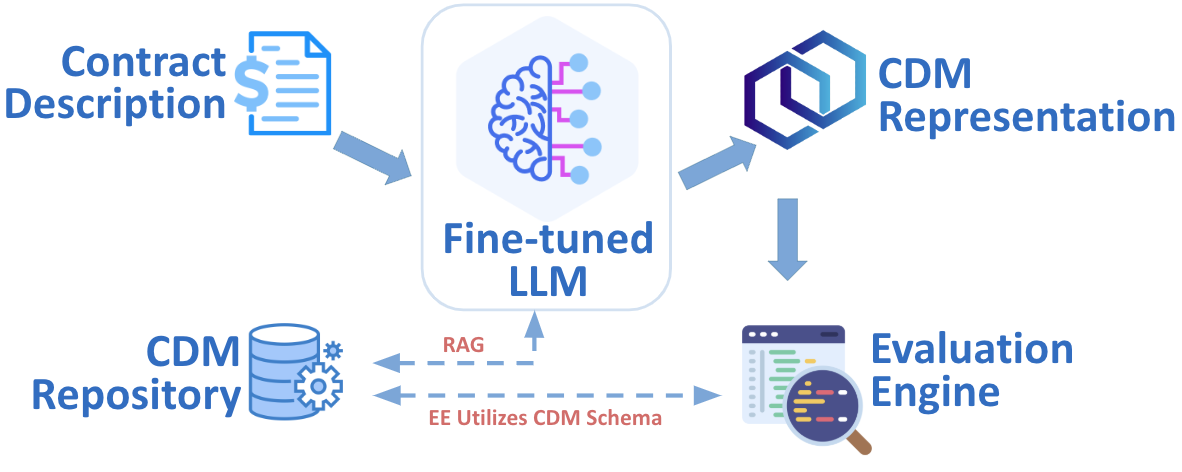}
    \caption{Workflow of directly generating a CDM representation from a given contract description using an LLM and RAG.}
    \label{fig:workflow_1}
\end{figure}

\begin{itemize}[topsep=0pt, partopsep=0pt, itemsep=0pt, parsep=0pt, left=0pt]
    \item \textbf{Contract Input}: The first step is to input the contract into the system. This is a raw contract description in natural language extracted from sources such as PDFs.
    
    \item \textbf{LLM Fine-tuning}: An LLM is fine-tuned using the previously discussed synthetic dataset consisting of contract descriptions paired with CDM examples. The fine-tuning leverages Parameter-Efficient Fine-Tuning (PEFT) techniques, including LoRA (Low-Rank Adaptation) with 4-bit quantization, to optimize specific parts of the model. This approach allows for efficient training while ensuring the model can accurately generate CDM representations from contract descriptions.

    \item \textbf{CDM Generation}: The contract is processed by a fine-tuned LLM to generate its CDM representation using a structured prompt specifically designed for contract-to-CDM conversion. To enhance accuracy and consistency, we integrate RAG. A structured knowledge base, built from relevant examples in the CDM repository, is preprocessed into retrievable chunks. The LLM retrieves the most relevant examples using semantic similarity and incorporates them into the input prompt, ensuring better alignment with CDM schema definitions. This combination of structured prompting and retrieved contextual knowledge guides the model in producing a well-formed CDM JSON representation. However, for experimental purposes, we evaluate both settings - CDM generation with and without RAG integration - to assess their comparative effectiveness. The evaluation procedure is discussed in detail in a later section of this paper.
   
\end{itemize}

\section{\emph{CDMizer}: A More Deterministic Approach for CDM Generation}

While the baseline approach provides an initial evaluation of LLM capabilities, it lacks determinism and may introduce inconsistencies in CDM representations. Moreover, due to the large size of the contracts and the potential for their CDM representations to span thousands of lines, the baseline method struggles to consistently generate complete JSON, hindered by token limitations and contract complexities.  To address these limitations, we propose \emph{CDMizer}, a more structured and deterministic method for converting contract descriptions into CDM representations. 

\emph{CDMizer} deterministically creates CDM templates and populates them using an LLM integrated with RAG. The CDM schema’s complexity, with deeply nested structures and interdependencies, makes direct generation from natural language unreliable, often leading to syntactic errors, schema non-compliance, or hallucinated fields. Templates enforce \textbf{100\%} schema adherence and correctness by providing a fixed structure, enabling component-wise generation where substructures are populated iteratively. By retaining only relevant fields, templates also mitigate token constraints and reduce complexity, ensuring a scalable and reliable CDM generation process.

Besides, existing methods~\cite{choudhury2018auto,tateishi2019automatic,marchesi2022automatic,allouche2021automatic,frantz2016institutions} for generating domain-specific code, especially smart contracts, show the effectiveness of the usage of templates for domain-specific tasks.

Given these advantages, we design minimal yet comprehensive templates for different contract types by leveraging the CDM schema and available examples. These templates are then populated with extracted contract details using LLMs, ensuring structured and more reliable CDM representations.
Note that these templates are designed solely to facilitate \emph{CDMizer}’s generation process and are not intended as a feature of the CDM.

\subsection{Template Creation}

The process of template creation involves several structured steps to ensure the relevance and usability of the resulting templates:

\paragraph{Schema Parsing:}
While CDM is published in many different formats and programming languages, the CDM JSON schema can be viewed as a collection of interconnected JSON files, each defining objects and their properties. Many properties reference other schema files via \texttt{\$ref}, forming a hierarchical structure. While comprehensive, parsing the entire schema generates an impractically large representation (e.g., exceeding four million lines of JSON). To manage this complexity, we focus on extracting only the fields necessary for the specific use case.

\paragraph{CDM Example Analysis:}
To identify relevant fields, we analyze example CDM representations of contracts of a specific type. These examples specify concrete instances of contract data and implicitly define the paths that should be preserved in the template.

\begin{algorithm}[h]
\footnotesize
\caption{Template Creation for CDM Representation}
\label{alg:template_creation}
\textbf{Input:} Root schema file $R$, Schema directory $D$, Example folder $E$ \\
\textbf{Output:} Pruned template $T$
\begin{algorithmic}[1]

\STATE \textbf{Step 1: Extract Relevant Keys}
\STATE Initialize $K \gets \emptyset$
\FOR{each example file $e \in E$}
    \STATE Load $e$ as a JSON object
    \STATE Flatten $e$ to extract dot-separated keys and add to $K$
\ENDFOR

\STATE \textbf{Step 2: Traverse Schema and Build Parse Tree}
\STATE $T \gets \textsc{TraverseSchema}(R, D, K, \texttt{""})$

\STATE \textbf{Subroutine: \textsc{TraverseSchema}}
\STATE Load schema $S$ from $R$
\STATE Initialize $T \gets \{\}$
\FOR{each property $p \in S.\texttt{properties}$}
    \STATE Update $path$: \\
    \hspace{10pt} \textbf{if} current path is empty \textbf{then} $path \gets p$ \\
    \hspace{10pt} \textbf{else} $path \gets \texttt{path} + \texttt{"."} + p$
    \IF{$path \notin K$}
        \STATE Skip $p$
    \ENDIF
    \IF{$p$ references another schema (\texttt{\$ref})}
        \STATE Resolve reference $ref$ as the file path for $p.\texttt{\$ref}$
        \STATE $T[p] \gets \textsc{TraverseSchema}(ref, D, K, path)$
    \ELSIF{$p$ is an array with a referenced schema}
        \STATE Resolve the reference $ref$ as the file path for $p.\texttt{items.\$ref}$
        \STATE $T[p] \gets [\textsc{TraverseSchema}(ref, D, K, path)]$
    \ELSE
        \STATE Add placeholder for $p$ to $T$
    \ENDIF
    \STATE Add $S.\texttt{description}$ to $T[p]$ if available
\ENDFOR
\RETURN $T$

\STATE \textbf{Output:} $T$

\end{algorithmic}
\end{algorithm}

\paragraph{Key Extraction:} A critical step involves flattening and extracting all keys from the examples into a dot-separated format. This enables efficient matching between example data and schema paths. Only fields traversed in at least one example are retained in the template.

\paragraph{Recursive Schema Traversal:} The schema is recursively traversed from the root, resolving properties that reference other schema (\texttt{\$ref}) and parsing the referenced structures. If a property path matches a key in the flattened example set, it is retained in the template, while unrelated properties are pruned. Additionally, field descriptions from the schema definition are included to provide semantic context.

\paragraph{Description Fields:} Each retained field in the template is annotated with a \texttt{description} key derived from the schema definition. It provides context about the field’s purpose and semantics and facilitates downstream tasks such as populating the template with natural language descriptions.
    
\paragraph{Placeholder Insertion:} Placeholder values are inserted for each field based on its data type: Strings- \texttt{""}, Arrays- \texttt{[]}, Objects- \texttt{\{\}} and Dates- \texttt{YYYY-MM-DD}.
These placeholders allow the template to be easily populated with actual data.

\paragraph{Pruning Irrelevant Fields:} After traversing the schema, fields not referenced in the examples are removed. Additionally, empty or unused structures are cleaned to produce a concise template.
    
The structured steps for template creation, as outlined above, are formally detailed in Algorithm~\ref{alg:template_creation}.

\begin{figure*}[t]
    \centering
    \includegraphics[width=0.8\linewidth]{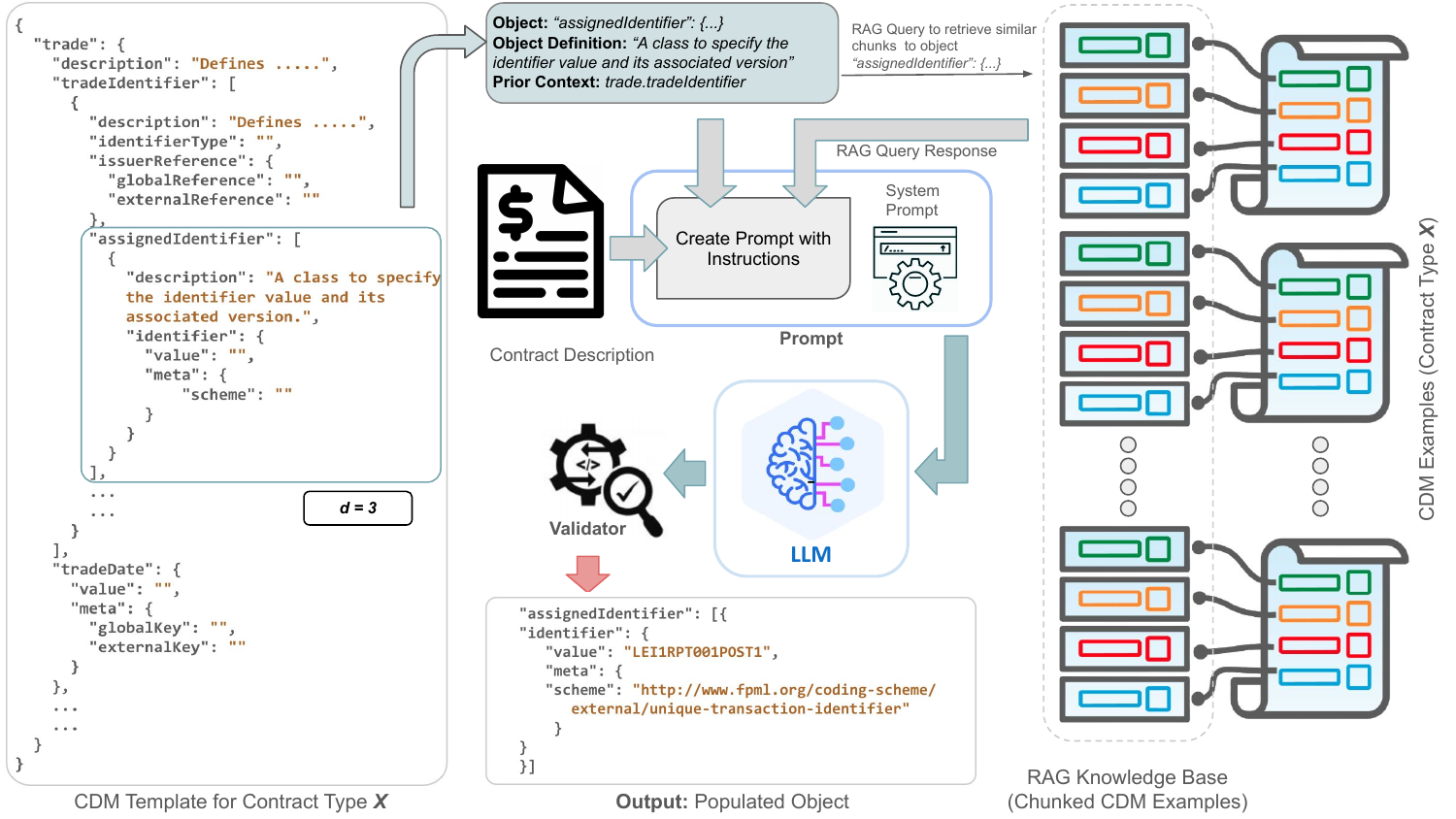}
    \caption{\emph{CDMizer} Workflow. Recursive traversal, governed by a depth threshold (\(d\)), selects substructures (e.g., \texttt{assignedIdentifier}) where the deepest subtree has a depth \(\leq d\). This ensures manageable task sizes for efficiency and accuracy. Context-aware prompts, incorporating object definitions, traversal paths, and RAG-retrieved examples, guide the LLM in populating fields, which are then validated. Recursive traversal ensures the entire structure is systematically completed.}
    \label{fig:workflow_2}
\end{figure*}

\subsection{Populating the Template}

Once a contract-specific template is defined, it is populated with relevant information extracted from the natural language contract description, forming its CDM representation. This process leverages template traversal, context-aware prompting, and Retrieval-Augmented Generation (RAG) to ensure semantic accuracy. The workflow is illustrated in Figure~\ref{fig:workflow_2}.

\subsubsection{Template Traversal and Depth Evaluation}
    The population process begins with a recursive traversal of the template’s JSON tree. During this traversal, the algorithm calculates the depth of the deepest subtree for each node.
    
    \paragraph{Depth Threshold (\(\boldsymbol{d}\)):} This parameter balances efficiency and accuracy by limiting processing to substructures with depth \(\leq d\), preventing token overflows and ensuring incremental processing of smaller, manageable substructures. Literature~\cite{karanjai2024solmover} suggests that they operate more effectively when presented with smaller, well-defined subtasks. Limiting the scope of the prompt to localized substructures improves the precision and relevance of the outputs. Treated as a hyperparameter, \(d\) is optimized through experimentation to achieve the best trade-off between accuracy and efficiency in generating semantically aligned CDM representations.

    \subsubsection{Constructing Context-Rich Prompts for Objects}
    For each selected object, the process involves constructing a detailed prompt to guide the LLM in generating the appropriate content. The prompt includes the following components:
    
    \begin{itemize}[topsep=0pt, partopsep=0pt, itemsep=0pt, parsep=0pt, left=0pt]
        \item \textbf{Current Object Structure:} The structure of the object being populated, including placeholders (e.g., \texttt{"assignedIdentifier": \{...\}}).
        \item \textbf{Object Definition:} A description extracted from the object’s schema (e.g., "A class to specify the identifier value and its associated version").
        \item \textbf{Traversal Context:} The hierarchical path leading to the current object (e.g., \texttt{trade.tradeIdentifier}), which situates the object within the broader template structure.
    \end{itemize}

    \subsubsection{Enhancing Prompts with RAG-Based Contextual Guidance}    
    To improve the LLM’s accuracy, the methodology integrates Retrieval-Augmented Generation (RAG), leveraging real-world examples as a knowledge base. First, examples of CDM representations for the given contract type are preprocessed and segmented into smaller chunks while preserving logical structures, forming a structured knowledge-base, as illustrated in Figure~\ref{fig:workflow_2}. During inference, a query is formulated based on the structure and description of the current object, retrieving semantically and structurally relevant chunks from the knowledge base. These retrieved examples are then incorporated into the prompt, providing contextual guidance to the LLM and ensuring that the generated content aligns more closely with established CDM representations.

    \vspace{-0.1em}
    \subsubsection{Prompt Submission and LLM Response}    
    The fully constructed prompt is submitted to the LLM, which generates the populated object. The prompt includes:
    
    \begin{itemize}[topsep=0pt, partopsep=0pt, itemsep=0pt, parsep=0pt, left=0pt]
        \item \textbf{Explicit Instructions:} Clear directives on how to populate the object, including required data types, formats, and contextual considerations.
        \item \textbf{Auxiliary Knowledge:} Examples retrieved through RAG illustrate expected field content and structure.
        \item \textbf{System Prompt:} Along with the constructed prompt, a system-level prompt is employed to establish general guidelines and expectations for the LLM, ensuring consistent outputs across various object populations.
    \end{itemize}
    
    As shown in Figure~\ref{fig:workflow_2}, the LLM uses this comprehensive prompt to populate fields like \texttt{identifier.value} and \texttt{identifier.value.meta.scheme} within the \texttt{assignedIdentifier} object.

    \subsubsection{Validation}
    After receiving the LLM’s response, the output undergoes a validation that ensures that the populated object has the exact same structure as the input object conforming to the schema.

    \subsubsection{Recursive Traversal and CDM Finalization}
    
    The traversal proceeds recursively, applying this process to each object having depth \(\leq d\) within the JSON tree. The recursive nature of the traversal ensures that all fields are processed systematically, maintaining hierarchical consistency. After populating all the objects, the template goes through a cleaning step, during which any empty field is removed. This cleaning step finally yields the CDM representation for the given natural language contract description.

\section{Experimental Evaluation}
To evaluate the effectiveness of the proposed approaches, we conducted experiments on a sample of 30 contracts from the synthetic dataset, with five contracts selected from each of six contract type categories. These 30 contracts were tested across four configurations: \textbf{Baseline w/o RAG}, \textbf{Baseline w/ RAG}, \textbf{\emph{CDMizer} w/o RAG}, and \textbf{\emph{CDMizer} w/ RAG}. However, fine-tuning was excluded from the baseline configurations due to critical limitations when applied to the contracts:

\begin{itemize}[topsep=0pt, partopsep=0pt, itemsep=0pt, parsep=0pt, left=0pt]
    \item \textbf{Token Limitations:} Contract descriptions and their CDM representations are lengthy, with fields such as \texttt{globalKey}, \texttt{meta}, and \texttt{identifiers} consuming significant token space without contributing meaningful information. This often led to truncation and incomplete outputs.

    \item \textbf{Strict Fine-Tuning Behavior:} Fine-tuned models rigidly mimicked dataset examples, generating metadata-heavy CDMs that prioritized non-critical fields. This exacerbated token limitations by leaving insufficient room for essential contract details.

    \item \textbf{Trade-offs Between Token Limits and Inference Time:} While increasing the token limit enabled more complete outputs, it resulted in prohibitively long inference times. Reducing the token limit caused incomplete representations, making fine-tuning impractical for evaluating longer contracts effectively.
\end{itemize}

\subsection{Evaluation Framework}

The generated CDM representations are assessed using predefined evaluation metrics:
    
        \paragraph{Syntactical Correctness}: Measures the proportion of generated keys that exist in the official CDM schema.
        \paragraph{Schema Adherence}: Evaluates whether the generated CDM structure conforms strictly to the CDM schema. The process involves traversing the generated JSON tree and comparing it with the CDM schema to identify which keys match and which do not.
        \paragraph{Semantic Coverage}: Determines how comprehensively the generated CDM captures the contract description’s key details while identifying irrelevant or missing information.

        

        We explored multiple approaches to assess semantic coverage, including direct LLM-based assessment, n-gram analysis, and cosine similarity between key terms. Initially, we considered direct LLM-based scoring, where the LLM was prompted to provide a holistic coverage score by comparing the contract description and the generated CDM. However, this approach proved unreliable, as the scores were often inconsistent and did not accurately reflect the degree of coverage.
        Next, we attempted n-gram analysis to extract key terms from the contract text by breaking it into token sequences. However, LLMs proved more effective at this task, as they can capture multi-word expressions and nuanced relationships that n-grams often miss.
        Cosine similarity was also considered for matching extracted key terms to CDM fields, but it often produced high scores for semantically unrelated terms, simply due to shared structures or numerical formats, leading to overestimated coverage, making this approach unsuitable for our evaluation framework.
        Ultimately, we found out that the most effective evaluation method was guiding an LLM by structuring its task as a step-by-step process that mimics how a human would systematically compute coverage. Instead of directly asking the LLM to estimate a coverage score, we designed a structured prompt that provides explicit step-by-step rules. These rules instruct the LLM to sequentially process the contract description, extract key details sentence by sentence, and compare them against the generated CDM. The LLM then categorizes information into three lists:

        \begin{itemize}[topsep=0pt, partopsep=0pt, itemsep=0pt, parsep=0pt, left=0pt]
            \item \textbf{Captured Information:} Contains correctly captured information in the CDM representation.
            \item \textbf{Uncaptured Information:} Contains information that is present in the contract description but not in the CDM representation.
            \item \textbf{Extraneous Information:} Additional content in the CDM representation that does not correspond to the contract description.
        \end{itemize}
        
        By explicitly defining the comparison process in the prompt and enforcing a structured methodology, we ensure that the LLM follows a deterministic approach, leading to more accurate and consistent evaluations of semantic coverage.

        Finally, we compute the \textbf{coverage score} using the following formula:

    \begin{equation}
        Coverage Score = \frac{C \times 100}{C + \mu \times U + \epsilon \times E}
    \end{equation}
    
    where \(C\), \(U\) and \(E\) represent the total number of \textbf{captured}, \textbf{uncaptured}, and \textbf{extraneous} elements, respectively.
    
    The weighting factors \(\mu\) and \(\epsilon\) are introduced to adjust for potential noise in the evaluation.

    \paragraph{\(\boldsymbol{\mu}\)}: Uncaptured elements may include details that are contextually relevant but not strictly contract-specific, as synthetic contract descriptions were generated using two real contract examples as references, and LLM might copy information from those. To adjust for this, we reduce the weight of \(U\) using the coefficient \(\mu\).

    \paragraph{\(\boldsymbol{\epsilon}\)}: Extraneous elements may result from fine-tuning and RAG-based retrieval, introducing information not explicitly present in the contract. To minimize their impact, we scale down the weight of \(E\) using the coefficient \(\epsilon\).
    
    These weighted adjustments ensure a more balanced coverage score by reducing the influence of non-essential or externally introduced elements.

\subsection{Results and Discussion}

\begin{table*}
    \centering
    \begin{tabular}{lcccccc}
        \toprule
        \multirow{3}{*}{Contract Type} & \multicolumn{3}{c}{Mean Syntactical Correctness (\%)} & \multicolumn{3}{c}{Mean Schema Adherence (\%)} \\
        \cmidrule(lr){2-4} \cmidrule(lr){5-7}
        & \shortstack{Baseline \\ w/o RAG}
         & \shortstack{Baseline \\ w/ RAG}
         & \multirow{-2}{*}{\emph{CDMizer}}
         & \shortstack{Baseline \\ w/o RAG}
         & \shortstack{Baseline \\ w/ RAG}
         & \multirow{-2}{*}{\emph{CDMizer}} \\
        \midrule
        Interest Rate Swap      
          & 46.80$_{(\pm28.01)}$ & 84.50$_{(\pm6.03)}$ & 100
          & 24.79$_{(\pm16.65)}$ & 85.44$_{(\pm1.67)}$ & 100 \\
        Equity Swap         
          & 54.99$_{(\pm24.41)}$ & 84.05$_{(\pm11.49)}$ & 100
          & 35.87$_{(\pm21.22)}$ & 84.81$_{(\pm11.89)}$ & 100 \\
        Equity Option       
          & 58.27$_{(\pm19.32)}$ & 94.45$_{(\pm2.95)}$ & 100
          & 43.14$_{(\pm25.92)}$ & 91.52$_{(\pm3.20)}$ & 100 \\
        Foreign Exchange Derivatives  
          & 64.61$_{(\pm7.52)}$ & 88.34$_{(\pm6.26)}$ & 100
          & 34.66$_{(\pm13.49)}$ & 87.67$_{(\pm6.38)}$ & 100 \\ 
        Commodity Option         
          & 60.54$_{(\pm11.28)}$ & 87.14$_{(\pm5.46)}$ & 100
          & 49.56$_{(\pm7.96)}$ & 93.02$_{(\pm3.59)}$ & 100 \\
        Credit Default Swap 
          & 63.82$_{(\pm10.78)}$ & 79.64$_{(\pm5.09)}$ & 100
          & 49.02$_{(\pm17.95)}$ & 80.59$_{(\pm7.25)}$ & 100 \\    
        \bottomrule
    \end{tabular}
    \caption{Comparison of mean syntactical correctness and schema adherence scores across different methods for each contract type (5 contracts for each type). Subscript values indicate \textbf{Standard Deviations}. As the approach suggests, \emph{CDMizer} (both with or without RAG versions) guarantees a score of 100\%.}
    \vspace{-0.6em}
    \label{tab:mean_scores}
\end{table*}

For generating the results, we set the depth threshold to \( d = 4 \), with weighting factors \( \mu = 0.3 \) and \( \epsilon = 0.1 \) to balance the impact of uncaptured and extraneous elements in the semantic coverage evaluation. 
These values were chosen after testing different configurations and comparing the resulting scores to human evaluations, ensuring the final setup captured contract details accurately while minimizing the impact of minor, contextually irrelevant elements.

As shown in Table \ref{tab:mean_scores}, \emph{CDMizer}, both with and without RAG, achieved 100\% syntactical correctness and schema adherence across all contract types. This outcome is expected due to its template-driven approach, which ensures strict compliance with the CDM schema. In contrast, \textit{Baseline w/o RAG} performed poorly, while \textit{Baseline w/ RAG} showed noticeable improvements but struggled with schema adherence, underscoring the challenges of generating structured JSON directly from contract descriptions.


\begin{figure}
    \centering
    \includegraphics[width=1\linewidth]{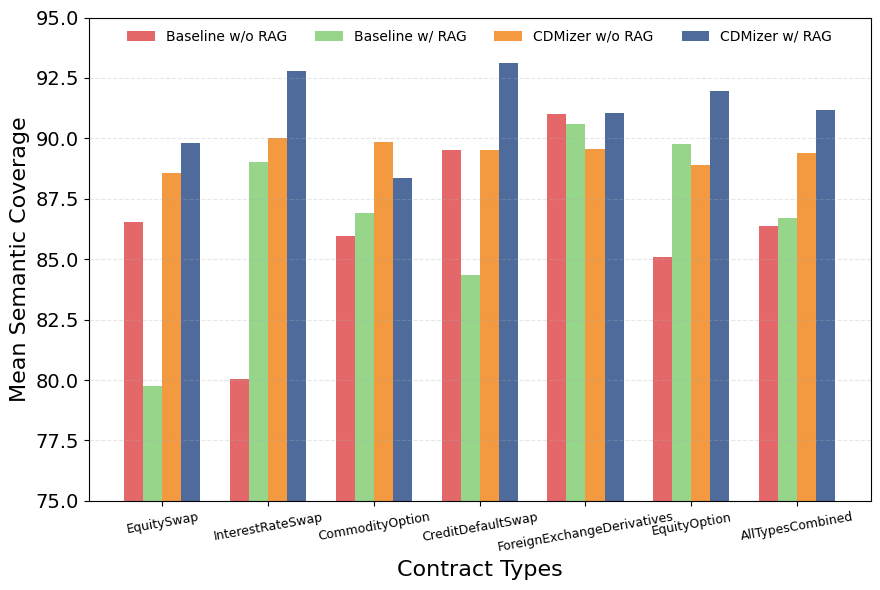}
    \caption{Mean semantic coverage scores for different methods across contract types. Highlights \emph{CDMizer}’s consistently improved performance over the baseline. This is more clearly visible (last set of bars) when the mean is calculated over the combined set of test contracts from all of the types.}
    \label{fig:contract_wise_bars}
    \vspace{-1em}
\end{figure}

Semantic coverage, illustrated in Figure \ref{fig:contract_wise_bars}, provides further insight into how effectively the generated CDM representations capture key contract details. In general, \emph{CDMizer} outperformed the baseline methods, and \textit{w/ RAG} methods achieved higher semantic coverage than \textit{w/o RAG} for both the baseline and \emph{CDMizer}, indicating the overall benefit of retrieval augmentation. However, there were instances where the \textit{w/o RAG} variant performed better than \textit{w/ RAG}—notably in \textit{Equity Swap} and \textit{Credit Default Swap} for the baseline method and in \textit{Commodity Option} derivatives for \emph{CDMizer}. This suggests that the LLM may have learned more about the CDM structure from the retrieved examples rather than contract-specific information in some cases. Additionally, RAG may introduce contextual information that was not explicitly stated in the contract, leading to the inclusion of extraneous details in the generated CDM. Despite these variations, when results were aggregated across all contract types, \emph{CDMizer} significantly outperformed the baseline methods, and retrieval augmentation generally proved beneficial.

These results were obtained using \textbf{Llama-3.1-8B-Instruct}, the best-performing LLM for consistency and accuracy. Table \ref{tab:llm_ranking} presents a comparison of different LLMs for this task. Since most LLMs failed to generate complete JSON representations in the baseline configurations, the LLM evaluation was conducted only on \emph{CDMizer}. As shown in the table, \textbf{Llama-3.1-8B-Instruct} achieved the highest semantic coverage, demonstrating its superior ability to handle complex schema while maintaining contextual accuracy.

\begin{table}[h]
    \centering
    \begin{tabular}{lcc}
        \toprule
        \multirow{3}{*}{LLM} & \multicolumn{2}{c}{\shortstack{Mean Semantic \\ Coverage (\%)}} \\
        \cmidrule(lr){2-3}
         & w/o RAG & w/ RAG \\
        \midrule
        Llama3.1-8B-Instruct      & 89.40 & 91.40 \\
        Deepseek-coder-6.7B-Instruct & 86.31 & 85.89 \\
        Mistral-7B-Instruct-v0.3  & 83.32 & 85.19 \\
        Llama3.2-3B-Instruct      & 84.13 & 84.70 \\
        \bottomrule
    \end{tabular}
    \caption{Mean Semantic Coverage overall test contracts across different LLMs with and without RAG for \emph{CDMizer}.}
    \label{tab:llm_ranking}
\end{table}

While our experiments primarily used synthetic contracts generated from CDM examples, we expect reasonable generalization to real contracts, given the structural consistency enforced by the CDM schema. However, real contracts often contain nuanced language, implicit cross-references, and complex conditional clauses, like the nested conditions and context-dependent triggers in, \textit{“The issuer may, at its option, for each period from and including the period starting 26 November 2009, upon giving 5 business days' notice, irrevocably switch the coupon of the notes to a fixed rate 3.00\% per annum.”} Capturing such dependencies requires a deeper understanding of temporal context and broader financial terms, which can lead to incomplete or contextually inaccurate mappings. Additionally, real contracts vary significantly in structure and terminology, creating challenges in aligning their content with predefined CDM templates. Despite these challenges, the template-driven nature of \emph{CDMizer} guarantees 100\% syntactical correctness and schema adherence, though it does not fully address potential semantic inaccuracies. Future work should include testing on a broader set of real contracts and may require fine-tuning, domain adaptation, or more advanced prompt engineering to handle these complexities.

\section{Conclusion and Future Directions}
This paper introduced a comprehensive framework for converting unstructured OTC financial derivative contracts into Common Domain Model (CDM) representations. We developed a baseline LLM-based pipeline, a deterministic template-driven approach (\emph{CDMizer}), and an LLM-based evaluation framework. \emph{CDMizer} ensured schema adherence and scalability, outperforming direct generation methods, especially with retrieval augmentation. Experimental evaluation demonstrated the effectiveness of structured generation and the impact of model selection, with Llama-3.1-8B-Instruct achieving the best performance. 
Challenges remain in capturing legal nuances, refining evaluations, and enabling enforceability. To address these limitations and further advance the automation of OTC contract processing, several avenues for future exploration are proposed:
(i) integrating \textbf{more features from the ISDA Master Agreement and legal documentation}~\cite{isda_ma} to capture non-operational aspects (e.g., bankruptcy, mergers) by applying CDMizer to ISDA contracts as covered by the legal agreement section of the CDM,
(ii) converting CDM into \textbf{executable smart contracts} for automated validation and enforceability, (iii) developing \textbf{more robust evaluation methods} beyond LLM-based validation, and (iv) leveraging \textbf{real-world contract examples} to critically assess the framework’s performance in complex legal contexts and iteratively enhance its robustness based on empirical insights.
By addressing these challenges and advancing these research directions, we move closer to a fully automated, reliable, and enforceable framework for OTC contract processing, bridging the gap between financial standardization and real-world applicability.


\section*{Acknowledgments}
{We acknowledge the support from NSF IUCRC CRAFT center research grant (CRAFT Grant \#22018) for this research. The opinions expressed in this publication do not necessarily represent the views of NSF IUCRC CRAFT. We are also grateful for the advice and resources from our CRAFT Industry Board members in shaping this work.}

\bibliographystyle{named}

\footnotesize{\bibliography{ijcai25}}

\begin{thebibliography}{}

\bibitem[\protect\citeauthoryear{Agreement}{2002}]{isda_ma}
ISDA~Master Agreement.
\newblock 2002 isda master agreement.
\newblock \url{https://www.isda.org/book/2002-isda-master-agreement-mylibrary/}, 2002.

\bibitem[\protect\citeauthoryear{Allamanis \bgroup \em et al.\egroup }{2024}]{allamanis2024unsupervised}
Miltiadis Allamanis, Sheena Panthaplackel, and Pengcheng Yin.
\newblock Unsupervised evaluation of code llms with round-trip correctness.
\newblock {\em arXiv preprint arXiv:2402.08699}, 2024.

\bibitem[\protect\citeauthoryear{Allouche \bgroup \em et al.\egroup }{2021}]{allouche2021automatic}
Mohamed Allouche, Mihai Mitrea, Alexandre Moreaux, and Sang-Kyun Kim.
\newblock Automatic smart contract generation for internet of media things.
\newblock {\em ICT Express}, 7(3):274--277, 2021.

\bibitem[\protect\citeauthoryear{Armitage}{2022}]{armitage2022trust}
Matthew Armitage.
\newblock Trust, confidence, and automation: The isda master agreement as a smart contract.
\newblock {\em Business Law Review}, 43(2), 2022.

\bibitem[\protect\citeauthoryear{Choudhury \bgroup \em et al.\egroup }{2018}]{choudhury2018auto}
Olivia Choudhury, Nolan Rudolph, Issa Sylla, Noor Fairoza, and Amar Das.
\newblock Auto-generation of smart contracts from domain-specific ontologies and semantic rules.
\newblock In {\em 2018 IEEE International Conference on Internet of Things (iThings) and IEEE Green Computing and Communications (GreenCom) and IEEE Cyber, Physical and Social Computing (CPSCom) and IEEE Smart Data (SmartData)}, pages 963--970. IEEE, 2018.

\bibitem[\protect\citeauthoryear{Clack and Vanca}{2018}]{clack2018temporal}
Christopher~D Clack and Gabriel Vanca.
\newblock Temporal aspects of smart contracts for financial derivatives.
\newblock In {\em Leveraging Applications of Formal Methods, Verification and Validation. Industrial Practice: 8th International Symposium, ISoLA 2018, Limassol, Cyprus, November 5-9, 2018, Proceedings, Part IV 8}, pages 339--355. Springer, 2018.

\bibitem[\protect\citeauthoryear{{FINOS}}{2024a}]{finos-cdm-repository}
{FINOS}.
\newblock Common domain model.
\newblock \url{https://github.com/finos/common-domain-model}, 2024.
\newblock Accessed: 2025-05-09.

\bibitem[\protect\citeauthoryear{FINOS}{2024b}]{cdm_purpose}
FINOS.
\newblock Overview of the finos cdm.
\newblock \url{https://cdm.finos.org/docs/cdm-overview/#purpose}, 2024.

\bibitem[\protect\citeauthoryear{FINOS}{2025}]{finos_cdm}
FINOS.
\newblock common-domain-model.
\newblock \url{https://cdm.finos.org/}, 2025.

\bibitem[\protect\citeauthoryear{Frantz and Nowostawski}{2016}]{frantz2016institutions}
Christopher~K Frantz and Mariusz Nowostawski.
\newblock From institutions to code: Towards automated generation of smart contracts.
\newblock In {\em 2016 IEEE 1st International Workshops on Foundations and Applications of Self* Systems (FAS* W)}, pages 210--215. IEEE, 2016.

\bibitem[\protect\citeauthoryear{Fries and Kohl-Landgraf}{2018}]{fries2018smart}
Christian~P Fries and Peter Kohl-Landgraf.
\newblock Smart derivative contracts (detaching transactions from counterparty credit risk: Specification, parametrisation, valuation).
\newblock {\em Available at SSRN 3163074}, 2018.

\bibitem[\protect\citeauthoryear{ISDA}{2023}]{isda}
ISDA.
\newblock Key trends in the size and composition of otc derivatives markets in the first half of 2023.
\newblock \url{https://www.isda.org/2023/12/07/key-trends-in-the-size-and-composition-of-otc-derivatives-markets-in-the-first-half-of-2023}, 2023.

\bibitem[\protect\citeauthoryear{{J.P. Morgan}}{2024}]{jpm-usd-inr-irs}
{J.P. Morgan}.
\newblock Usd/inr irs disclosure.
\newblock \url{https://www.jpmorgan.com/content/dam/jpm/global/disclosures/IN/usd-inr-irs.pdf}, 2024.
\newblock Accessed: 2025-05-09.

\bibitem[\protect\citeauthoryear{Kang \bgroup \em et al.\egroup }{2024}]{kang2024using}
Inwon Kang, William~Van Woensel, and Oshani Seneviratne.
\newblock Using large language models for generating smart contracts for health insurance from textual policies.
\newblock In {\em AI for Health Equity and Fairness: Leveraging AI to Address Social Determinants of Health}, pages 129--146. Springer, 2024.

\bibitem[\protect\citeauthoryear{Karanjai \bgroup \em et al.\egroup }{2024}]{karanjai2024solmover}
Rabimba Karanjai, Lei Xudagger, and Weidong Shi.
\newblock Solmover: Feasibility of using llms for translating smart contracts.
\newblock In {\em 2024 IEEE International Conference on Blockchain and Cryptocurrency (ICBC)}, pages 1--3. IEEE, 2024.

\bibitem[\protect\citeauthoryear{Li \bgroup \em et al.\egroup }{2022}]{li2022competition}
Yujia Li, David Choi, Junyoung Chung, Nate Kushman, Julian Schrittwieser, R{\'e}mi Leblond, Tom Eccles, James Keeling, Felix Gimeno, Agustin Dal~Lago, et~al.
\newblock Competition-level code generation with alphacode.
\newblock {\em Science}, 378(6624):1092--1097, 2022.

\bibitem[\protect\citeauthoryear{Liu \bgroup \em et al.\egroup }{2024}]{liu2024your}
Jiawei Liu, Chunqiu~Steven Xia, Yuyao Wang, and Lingming Zhang.
\newblock Is your code generated by chatgpt really correct? rigorous evaluation of large language models for code generation.
\newblock {\em Advances in Neural Information Processing Systems}, 36, 2024.

\bibitem[\protect\citeauthoryear{Marchesi \bgroup \em et al.\egroup }{2022}]{marchesi2022automatic}
Lodovica Marchesi, Katiuscia Mannaro, Michele Marchesi, and Roberto Tonelli.
\newblock Automatic generation of ethereum-based smart contracts for agri-food traceability system.
\newblock {\em Ieee Access}, 10:50363--50383, 2022.

\bibitem[\protect\citeauthoryear{Oluwajebe \bgroup \em et al.\egroup }{2020}]{oluwajebe2020smart}
Olusegun Oluwajebe, Mary Duah, and Polina Golnikova.
\newblock Smart derivatives contracting: Automating interest rate swaps in the over-the-counter (otc) market with the daml.
\newblock {\em Available at SSRN 3750089}, 2020.

\bibitem[\protect\citeauthoryear{Parvez \bgroup \em et al.\egroup }{2021}]{parvez2021retrieval}
Md~Rizwan Parvez, Wasi~Uddin Ahmad, Saikat Chakraborty, Baishakhi Ray, and Kai-Wei Chang.
\newblock Retrieval augmented code generation and summarization.
\newblock {\em arXiv preprint arXiv:2108.11601}, 2021.

\bibitem[\protect\citeauthoryear{{RBC Capital Markets}}{2024}]{rbc-structuredrates-2024}
{RBC Capital Markets}.
\newblock Structured rates presentation or document.
\newblock \url{https://www.rbccm.com/structuredrates/file-566984.pdf}, 2024.
\newblock Accessed: 2025-05-09.

\bibitem[\protect\citeauthoryear{Singh \bgroup \em et al.\egroup }{2024}]{singh2024option}
Srisht~Fateh Singh, Panagiotis Michalopoulos, and Andreas Veneris.
\newblock Option contracts in the defi ecosystem: Motivation, solutions, \& technical challenges.
\newblock In {\em 2024 IEEE International Conference on Blockchain and Cryptocurrency (ICBC)}, pages 1--7. IEEE, 2024.

\bibitem[\protect\citeauthoryear{Sorensen}{2024}]{sorensen2024correct}
Derek Sorensen.
\newblock (in) correct smart contract specifications.
\newblock In {\em 2024 IEEE International Conference on Blockchain and Cryptocurrency (ICBC)}, pages 567--575. IEEE, 2024.

\bibitem[\protect\citeauthoryear{Tateishi \bgroup \em et al.\egroup }{2019}]{tateishi2019automatic}
Takaaki Tateishi, Sachiko Yoshihama, Naoto Sato, and Shin Saito.
\newblock Automatic smart contract generation using controlled natural language and template.
\newblock {\em IBM Journal of Research and Development}, 63(2/3):6--1, 2019.

\bibitem[\protect\citeauthoryear{Vaithilingam \bgroup \em et al.\egroup }{2022}]{vaithilingam2022expectation}
Priyan Vaithilingam, Tianyi Zhang, and Elena~L Glassman.
\newblock Expectation vs. experience: Evaluating the usability of code generation tools powered by large language models.
\newblock In {\em Chi conference on human factors in computing systems extended abstracts}, pages 1--7, 2022.

\bibitem[\protect\citeauthoryear{Van~Woensel \bgroup \em et al.\egroup }{2023}]{van2023translating}
William Van~Woensel, Manan Shukla, and Oshani Seneviratne.
\newblock Translating clinical decision logic within knowledge graphs to smart contracts.
\newblock In {\em SeWeBMeDA@ ESWC}, 2023.

\end{thebibliography}

\end{document}